\documentclass[twocolumn,aps,prd,showpacs,preprintnumbers,nofootinbib]{revtex4-1}
\usepackage{amsfonts,amssymb,amsmath,hyperref}
\newcommand{\f}[2]{\frac{#1}{#2}}
\newcommand{\mk}[1]{\left( #1 \right)}
\newcommand{\kk}[1]{\left[ #1 \right]}

\newcommand{\be}{\begin{equation}}
\newcommand{\ee}{\end{equation}}
\def\p{\partial}
\def\vx{\mbox{\boldmath $x$} }
\def\vy{\mbox{\boldmath $y$} }
\def\vp{\mbox{\boldmath $p$} }
\def\vq{\mbox{\boldmath $q$} }
\def\vQ{\mbox{\boldmath $Q$} }
\def\vP{\mbox{\boldmath $P$} }
\begin{document}%%%%%%%%%%%%%%%%%%%%%%%%%%%%%%%%%%%%%%%%%
\preprint{RESCEU-48/14}

\title{
Third order equations of motion and the Ostrogradsky instability
}

\author{Hayato Motohashi}
%\email[E-mail:]{motohashi"at"kicp.uchicago.edu}
\affiliation{Kavli Institute for Cosmological Physics, The University of Chicago, 
Chicago, Illinois 60637, U.S.A.
}

\author{Teruaki Suyama}
%\email[E-mail:]{suyama"at"resceu.s.u-tokyo.ac.jp}
\affiliation{Research Center for the Early Universe (RESCEU),
Graduate School of Science,
The University of Tokyo, Tokyo 113-0033, Japan
}

\begin{abstract}%%%%%%%%%%%%%%%%%%%%%%%%%%%%%%%%%%%%%%%%%
It is known that any nondegenerate Lagrangian containing time 
derivative terms higher than first order suffers from the Ostrogradsky
instability, pathological excitation of positive and negative energy degrees of freedom.
We show that, within the framework of analytical mechanics of point particles, 
any Lagrangian for third order equations of motion, which evades
the nondegeneracy condition, still leads to the Ostrogradsky instability.
Extension to the case of higher odd order equations of motion is also considered.
\end{abstract}
\pacs{}
\maketitle
%%%%%%%%%%%%%%%%%%%%%%%%%%%%%%%%%%%%%%%%%

\section{Introduction}%%%%%%%%%%%%%%%%%%%%%%%%%%%%%%%%%%%%%%%%%
\label{sec:I}
Nature prefers to describe the laws of physics by the second-order differential equations.
Newton's equation of motion (EOM),
Maxwell equations and Einstein equations etc.\ are all the second-order 
differential equations for positions of particles or fields.
Why the laws of physics are like this is a fundamental question of physics.

It is known that any nondegenerate Lagrangian containing time 
derivative terms higher than the first order yields a Hamiltonian which is not
bounded from below~\cite{Woodard:2006nt}.
This means that the energy of the system can take an arbitrarily negative value.
While such a system for free theory is not pathological, 
when it is coupled to a normal system with positive energy, the total system will quickly develop into excitation of positive and negative 
energy degrees of freedom.\footnote{For a possible loophole, see \cite{Smilga:2008pr}.}
This instability, recently called the Ostrogradsky instability,\footnote{Although it has been commonly referred to as ``Ostrogradsky instability'' recently and we will follow the convention in the present paper, it is not clear if Ostrogradsky himself made this statement in his original paper~\cite{Ostrogradsky}.
Furthermore, we are also not sure if the unboundedness of the Hamiltonian for higher derivative
Lagrangians, although it is referred to as ``theorem of Ostrogradsky'' in~\cite{Woodard:2006nt},
was shown by Ostrogradsky in Ref.~\cite{Ostrogradsky}.
As far as we know, Ref.~\cite{Pais:1950za} by Pais and Uhlenbeck is
one of the oldest work in which unboundedness of the Hamiltonian for a special
type of higher derivative Lagrangian was shown explicitly.}
is quite general and discussed in detail in~\cite{Woodard:2006nt}.
Here we briefly explain the main point of the Ostrogradsky instability
by following~\cite{Woodard:2006nt}.
In order to have the essence, we focus on a Lagrangian of 
$N$ variables
given by $L(\vx,{\dot \vx},{\ddot \vx})$,
where $\vx =(x_1,x_2,\cdots,x_N)$.
The Euler-Lagrange equations are given by
\begin{equation}
\frac{d^2}{dt^2}\frac{\partial L}{\partial {\ddot \vx}}-
\frac{d}{dt}\frac{\partial L}{\partial {\dot \vx}}+
\frac{\partial L}{\partial \vx}=0. \label{Euler-Lag}
\end{equation}
We require nondegeneracy, {\it i.e.} 
$\det \left( {\partial^2 L}/{\partial {\ddot x_i}\partial {\ddot x_j}} \right) \neq 0$.
Then the equations of motion are the fourth-order differential equations
and the number of the degrees of freedom as the number of independent initial 
conditions is $4N$.\footnote{There are literatures where the number of the degrees of freedom denotes the half of the number of the initial conditions, but this notation assumes the second-order EOM. Here, we denote that the number of the degrees of freedom as the number of the initial conditions.}
In order to move to the canonical formalism,
we need to define $4N$ canonical coordinates. 
They are given by~\cite{Ostrogradsky,Whittaker:1937lec},
\begin{equation}
\vQ_1=\vx,~~\vQ_2={\dot \vx},~~
\vP_1=\frac{\partial L}{\partial {\dot \vx}}-\frac{d}{dt} \frac{\partial L}{\partial {\ddot \vx}},~~
\vP_2=\frac{\partial L}{\partial {\ddot \vx}}.
\end{equation}
With the assumption of nondegeneracy,
we can solve the above equations for $\{\vx,{\dot \vx},{\ddot \vx},\vx^{(3)}\}$.
Then the Hamiltonian is defined in the standard manner,
\begin{equation}
H=\vP_1 \cdot {\dot \vQ_1}+\vP_2 \cdot {\dot \vQ_2}-L.
\end{equation}
We can confirm that  
the Hamilton's equations are equivalent to
the Euler-Lagrange equation.
Thus the Hamiltonian represents the generator of the time evolution
and can be interpreted as the energy of the system.
It is immediate to understand that ${\ddot \vx}$ is written in terms of
$\{\vQ_1,\vQ_2,\vP_2\}$.
This means that $\vP_1$ appears only in the first term and the Hamiltonian depends
on $\vP_1$ linearly.
Since there is no constraint among each element of $\vP_1$,
they can independently take any value.
Thus, the Hamiltonian can take an arbitrarily positive value or negative value, 
leading to the Ostrogradsky instability.

As the above argument shows,
the Ostrogradsky instability is a quite generic feature of higher derivative theories
and answers why nature is described by the second-order EOMs.
Due to its simplicity and generality,
the Ostrogradsky instability has also played a powerful role in constructing a sensible scalar-tensor theory
(for instance, \cite{Nicolis:2008in,Horndeski:1974wa,Deffayet:2011gz,Kobayashi:2011nu}),
mainly for the purpose of modifying gravity at infrared to explain the accelerated expansion of the Universe.

Now, it is intriguing to ask if the Ostrogradsky instability still persists 
when the EOMs (\ref{Euler-Lag}) are third-order differential equations.
Since the nondegenerate Lagrangian always yields the fourth-order EOMs,
the third-order EOMs must be obtained from the degenerate Lagrangian.
More generally, any nondegenerate Lagrangian yields the even order equations
while the odd order EOMs are obtained from a degenerate Lagrangian.
Clearly, the existence of the Ostrogradsky instability in the latter case 
is not covered by the argument provided above.
As far as we know, it has not been discussed in the literature whether 
the third-order EOMs (and higher odd order EOMs) 
are inevitably associated with the Ostrogradsky instability or not.
The aim of this paper is to answer this question.

Finally, it is important to make a distinction between 
the degeneracy of Lagrangian and the third-order EOMs.
Third-order EOMs are always derived from a degenerate Lagrangian but 
the opposite statement is not necessarily true (see also \cite{Chen:2012au} for relevant discussion).
For instance, although a Lagrangian of a single variable ($N=1$) given by $L={\ddot x}f({\dot x},x)$ is degenerate, its EOM is the second-order differential equation and the Ostrogradsky instability does not appear.
Thus, a statement that any degenerate Lagrangian suffers from the Ostrogradsky
instability is incorrect.
But a statement that any degenerate Lagrangian giving third-order EOMs suffers from  
the Ostrogradsky instability is correct, as we shall show below.

This paper is organized as follows.
In \S\ref{sec:L}, we construct the most general Lagrangian for the third-order EOMs 
in analytical mechanics of point particles.
In \S\ref{sec:H}, we show that the Ostrogradsky instability always
appears for the theory by performing the Hamiltonian analysis.
Extension to the case of higher odd order EOMs is also considered in the end of \S\ref{sec:H}, 
and \S\ref{sec:C} is devoted to conclusion.
Extension to the field theory is beyond the scope of this paper, although it has
interesting applications, for instance, like
the Chern-Simons gravity~\cite{Jackiw:2003pm} for which the gravitational field equations are third-order differential equations.

\section{Lagrangian for third-order EOMs}%%%%%%%%%%%%%%%%%%%%%%%%%%%%%%%%%%%%%%%%%
\label{sec:L}
Let us try to construct the most general Lagrangian that yields the third-order EOMs.
We will address general odd order EOMs after completing an analysis for third-order EOMs.
We again consider a Lagrangian which depends on $N$ variables $\vx=(x_1,\cdots,x_N)$, 
and contains up to $n$th time derivative:
\be \label{Lag-n} L=L(\vx,{\dot \vx},\vx^{(2)},\cdots,\vx^{(n)}). \ee
At this moment, $n \geq 2$ is an arbitrary positive integer.

We determine $n$ as well as the form of the Lagrangian by requiring that $L$ yields the third-order EOMs for $\vx$.
Since our primary concern is the Ostrogradsky instability of variables having 
the third-order EOMs, we require that all the variables obey the third-order EOMs
independently.
We do not consider their couplings to variables obeying at most the second-order EOMs.
Such coupling does not affect the existence of the Ostrogradsky instability.
The Euler-Lagrange equation for \eqref{Lag-n} is given by
\be \label{ELeq-single}
\frac{d^n}{dt^n} \frac{\p L}{\p \vx^{(n)}}
-\frac{d^{n-1}}{dt^{n-1}}\frac{\p L}{\p \vx^{(n-1)}}
+\frac{d^{n-2}}{dt^{n-2}}\frac{\p L}{\p \vx^{(n-2)}}-\cdots =0.
\ee
For a generic Lagrangian, the highest order derivative is $\vx^{(2n)}$,
that comes only from the first term of Eq.~\eqref{ELeq-single}.
For EOMs \eqref{ELeq-single} not to have $\vx^{(2n)}$, we require that 
\be \frac{\p^2 L}{\p x_i^{(n)}\p x_j^{(n)}} = 0,
\ee
for any $i,j=1,\cdots ,N$.
Thus the Lagrangian should be written as
\be
L=\sum_{j=1}^N x_j^{(n)} f_j(\vx,\cdots,\vx^{(n-1)})+
g(\vx,\cdots,\vx^{(n-1)}),
\ee
with arbitrary functions $f_j$ and $g$.
The Euler-Lagrange equation then reads 
\be
\sum_{j=1}^N x_j^{(2n-1)} \mk{ \f{\p f_i}{\p x_j^{(n-1)}} - \f{\p f_j}{\p x_i^{(n-1)}} } + (\text{lower deriv.}) = 0.
\ee
To obtain the third-order EOMs, we impose that the coefficient for $x_j^{(2n-1)}$ vanishes:
\be \f{\p f_i}{\p x_j^{(n-1)}} = \f{\p f_j}{\p x_i^{(n-1)}}. \ee
By using the Green's theorem, there exists $F$ that satisfies 
\be f_i= \f{\p}{\p x_i^{(n-1)}} F(\vx,\cdots, \vx^{(n-1)}). \ee
Therefore we can rewrite the Lagrangian as
\be L=-\sum_{j=1}^N\sum_{k=1}^{n-1} x_j^{(k)} \f{\p F}{\p x_j^{(k-1)}} + g, \ee
up to a difference of a total derivative $dF/dt$.
This means that, without a loss of generality,
we could have started from a Lagrangian containing at most $\vx^{(n-1)}$ in Eq.~(\ref{Lag-n}).
By repeating the same procedure, we can reduce the original Lagrangian (\ref{Lag-n})
to the one that contains at most $\ddot \vx$. 
From the absence of fourth-order derivative in the EOMs, 
the Lagrangian should be written as
\be \label{Lag-N} L=\sum_{j=1}^N \ddot x_j f_j(\vx,\dot \vx)+g(\vx,\dot \vx). \ee
The EOM is then given by
\be \sum_{j=1}^N M^{(\dot x)}_{ij} \dddot x_j +\text{(lower derivatives.)}=0, \ee
where we have defined
\be M^{(X)}_{ij} \equiv \f{\p f_i}{\p X_j} - \f{\p f_j}{\p X_i}. \ee
Since $M^{(\dot x)}$ is an $N\times N$ antisymmetric matrix, 
we can block-diagonalize it to arrive at
\be 
\Lambda=
\begin{pmatrix}
\Lambda_1 & 0 & \cdots \\
0 & \Lambda_2 & \cdots \\
\vdots & \vdots & \ddots 
\end{pmatrix}
\quad 
\text{with}
\quad
\Lambda_a=
\begin{pmatrix}
0 & \lambda_a  \\
-\lambda_a & 0 
\end{pmatrix}
.	
\ee
If $N$ is a odd number, $N=2I-1$ with $I\geq 1$, we have $\det \Lambda =0$ due to the Jacobi's theorem.
Thus, there exists a $J$ such that, $1\leq J\leq I-1$ and  
$(2J-1)$ lows and columns of $\Lambda$ are occupied by zero. 
Hence, not all the variables obey the third-order differential equations independently.
In other words, the degrees of freedom as the number of independent initial conditions 
is smaller than $3N$ and we can introduce a new set of $2(I-J)$ variables obeying
the third-order EOMs independently
and a new set of $2J$ variables obeying at most the second-order EOMs.
As we stated before, couplings to variables obeying the second-order EOMs are not essential
for the existence of the Ostrogradsky instability.
Therefore, we only need to consider the even $N$ case.
For even $N$, if $M^{(\dot x)}$ has vanishing $\lambda_a$, 
we can further reduce the number of variables and make all the $\lambda_a$ are nonvanishing.
Therefore, without loss of generality, 
we can set $N$ to be even in \eqref{Lag-N} and all the $\lambda_a$ are nonvanishing, 
{\it i.e.} $\det M^{(\dot x)}\neq 0$.

In conclusion, the most general Lagrangian that yields independent third-order EOMs for all the variables is the Lagrangian \eqref{Lag-N} for even number variables:
\be \label{L-3rd} L=\sum_{j=1}^{2N}  {\ddot x_j} f_j(\vx,{\dot \vx})+g(\vx,{\dot \vx})  , \ee
with $\det M^{(\dot x)}_{ij}\neq 0$.
Here, we have changed the notation of the number of variables from $N$ to $2N$.
Thus, $N$ is still an arbitrary positive integer.
With this setup,
%all the variables obey the third-order EOMs independently and 
we need $6N$ initial 
conditions for $x_i$, $\dot x_i$, $\ddot x_i$ for $i=1,\cdots,2N$ 
at given initial time $t=t_{\rm ini}$ to determine the time evolution of the system.

\section{Hamiltonian analysis}%%%%%%%%%%%%%%%%%%%%%%%%%%%%%%%%%%%%%%%%%
\label{sec:H}
We proceed to the Hamiltonian analysis of the most general 
Lagrangian for the third-order EOMs \eqref{L-3rd}.
Among several different approaches for constructing the canonical formalism 
for higher derivative theories~\cite{Plyushchay:1987eu,Nesterenko:1987jt,Batlle:1987ej,Saito:1989bt,Pons:1988tj,Andrzejewski:2010kz},
we adopt the one used in~\cite{Plyushchay:1987eu,Pons:1988tj},
which reduces the higher derivative Lagrangian to the standard one with constraints
by introducing the Lagrange multipliers. 
The advantage of this method is that we can use the standard Dirac's formalism
for the canonical formalism with constraints.
With the Lagrange multiplier $\lambda_i$, the Lagrangian \eqref{L-3rd} can be rewritten as 
\be \label{cL} L=\sum_{j=1}^{2N}\kk{ \dot y_j f_j(\vx,\vy) + \lambda_j(\dot x_j-y_j)}+g(\vx,\vy). \ee
Variation with respect to $\lambda_i$ yields $\dot x_i=y_i$, 
with which \eqref{cL} reproduces the original Lagrangian \eqref{L-3rd}.

Now the Lagrangian \eqref{cL} is written up to the first-order derivatives of $x_i$, $y_i$, $\lambda_i$. 
This implies that its EOMs are the second-order differential equations and 
we need $12N$ initial conditions in total. 
We will see that $6N$ initial conditions out of $12N$ initial conditions are 
fixed by second-class constraints and we are finally left with $6N$ initial conditions, 
which matches the number of initial conditions needed for the original Lagrangian \eqref{L-3rd}.

The canonical momentum for $x_i$, $y_i$, $\lambda_i$ is given by
\begin{align}
p_{xi}&\equiv \f{\p L}{\p \dot x_i}=\lambda_i,\\
p_{yi}&\equiv \f{\p L}{\p \dot y_i}=f_i(\vx,\vy),\\
p_{\lambda i}&\equiv \f{\p L}{\p \dot \lambda_i}=0.
\end{align}
Since all the momenta do not contain $\dot x$, $\dot y$, $\dot \lambda$, we cannot solve them in terms of momenta. This implies that we have $6N$ primary constraints:
\begin{align}
\phi_{xi}&\equiv p_{xi}-\lambda_i=0,\\
\phi_{yi}&\equiv p_{yi}-f_i(\vx,\vy)=0,\\
\phi_{\lambda i}&\equiv p_{\lambda i}=0.
\end{align}
Following the Hamiltonian formalism for constrained systems~\cite{Dirac:1950pj,Dirac:1958sq,Dirac:1964lec},
we incorporate these constraints by adding them into the Lagrangian with Lagrange multipliers.
Thus we consider the variation of the action
\be \label{S-con} S = \int^{t_2}_{t_1} dt \kk{ \sum_{j=1}^{2N} \sum_{q=x,y,\lambda}  (p_{qj}\dot q_j -\mu_{qj}\phi_{qj}) - H }, \ee
where the Hamiltonian $H$ is given by
\begin{align} 
H(\vp,\vq)&=\sum_{j=1}^{2N} \sum_{q=x,y,\lambda}  p_{qj}\dot q_j-L, \notag \\
&=\sum_{j=1}^{2N}\lambda_j y_j-g(\vx,\vy), \label{Ham}
\end{align}
and $\mu_{xi}$, $\mu_{yi}$, $\mu_{\lambda i}$ are the Lagrange multipliers for the primary constraints.
The variation of \eqref{S-con} yields the canonical equation
\begin{align}
\label{cano-q} \dot q_i&=\f{\p H}{\p p_{qi}} + \sum_{j=1}^{2N}\sum_{q=x,y,\lambda} \mu_{qj}\f{\p \phi_{qj}}{\p p_{qi}},\\
\label{cano-p}\dot p_{qi}&=-\f{\p H}{\p q_i} - \sum_{j=1}^{2N}\sum_{q=x,y,\lambda} \mu_{qj}\f{\p \phi_{qj}}{\p q_{i}}.
\end{align}
Therefore, the time evolution of any function $\xi(p,q)$ is governed by
\begin{align} 
\f{d\xi}{dt}&= \sum_{j=1}^{2N}\sum_{q=x,y,\lambda} \mk{ \f{\p \xi}{\p q_j}\dot q_j+\f{\p \xi}{\p p_{qj}}\dot p_{qj} }, \notag \\
&= %\f{\p \psi}{\p t}+
\{\xi, H\}+\sum_{j=1}^{2N} \sum_{q=x,y,\lambda}\mu_{qj}\{\xi,\phi_{qj}\}, \notag\\
&\approx \{ \xi, H_T \},
\end{align}
where the total Hamiltonian $H_T$ is given by
\be \label{HT} H_T=H+\sum_{j=1}^{2N} \sum_{q=x,y,\lambda} \mu_{qj}\phi_{qj}, \ee
and $\{ \ \}$ denotes the Poisson bracket 
\be \{ \xi, \eta \} = \sum_{j=1}^{2N} \sum_{q=x,y,\lambda} \mk{ \f{\p \xi}{\p q_j}\f{\p \eta}{\p p_{qj}} - \f{\p \xi}{\p p_{qj}}\f{\p \eta}{\p q_j} }. 
\ee
The weak equality $\approx$ expresses an equality that holds after performing the Poisson bracket and then imposing that all the constraints vanishes.

To make the constraints satisfied through the time evolution, we impose a consistency condition
\be \label{consis} \f{d\phi_{qi}}{dt}=\{ \phi_{qi}, H \} + \sum_{j=1}^{2N} \sum_{q=x,y,\lambda} \mu_{qi} \{ \phi_{qi}, \phi_{qj} \} \approx 0. \ee
For $q=x,y,\lambda$, this reads 
\begin{align}
\f{d\phi_{xi}}{dt}
&=\sum_{j=1}^{2N}\mu_{yi}\f{\p f_j}{\p x_i}-\mu_{\lambda i} + \f{\p g}{\p x_i},\\
\f{d\phi_{yi}}{dt}
&=-\lambda_i-\sum_{j=1}^{2N}\mk{\mu_{xj}\f{\p f_i}{\p x_j}+\mu_{yj} M^{(y)}_{ij}} - \f{\p g}{\p y_i},\\
\f{d\phi_{\lambda i}}{dt}
&=-y_i+\mu_{xi}.
\end{align}
Since $\det M^{(y)} \neq 0$ by definition,
there exists an inverse matrix $M_{(y)}^{-1}$.
Thus we can solve the consistency condition to determine all the Lagrangian multipliers
\begin{align}
\mu_{xi}=& y_i,\\
\mu_{yi}=& -\sum_{j=1}^{2N}(M_{(y)}^{-1})_{ij}\mk{\lambda_j+\f{\p g}{\p y_j}+\sum_{k=1}^{2N}y_k\f{\p f_j}{\p x_k}},\\
\mu_{\lambda i}=& -\sum_{j=1}^{2N}\sum_{k=1}^{2N}(M_{(y)}^{-1})_{jk}\mk{\lambda_k+\f{\p g}{\p y_k}
+\sum_{\ell=1}^{2N}y_\ell \f{\p f_k}{\p x_\ell}} \f{\p f_j}{\p x_i} \nonumber \\
&+\f{\p g}{\p y_i},
\end{align} 
which means that we have exhausted all the constraints and there is no additional (secondary) constraint.

Since all the Lagrange multipliers are determined, \eqref{consis} should be invertible, {\it i.e.} 
$\det \Delta \neq 0$ where $\Delta_{qiq'j}\equiv \{ \phi_{qi}, \phi_{q'j} \}$.
Indeed, the Poisson brackets between constraints are
\begin{align}
\{ \phi_{yi}, \phi_{yj} \} &= M^{(y)}_{ji},\\
\{ \phi_{xi}, \phi_{yj} \} &= \f{\p f_j}{\p x_i},\\
\{ \phi_{\lambda i}, \phi_{x j} \} &= \delta_{ij},
\end{align}
and all the other combinations vanish. 
Thus $\det \Delta=\det M^{(y)} \neq 0$, which is exactly the condition that allows us to solve for the Lagrange multipliers.
Therefore, all $6N$ constraints are second class, and removes $6N$ initial conditions.
Consequently, we are left with $12N-6N=6N$ initial conditions, as expected.

With these Lagrange multipliers, 
we obtain the self-consistent total Hamiltonian $H_T$ by \eqref{HT}, with 
which all the constraints are satisfied all the time if they are satisfied initially.
Considering \eqref{HT} on the constraint surface, 
we note that $H_T\approx H$ and \eqref{Ham} suggests that its first term is linear in
$p_{xi}$, which is not restricted by any primary constraints and can take an arbitrary value dynamically.
Therefore, the Hamiltonian can vary from $-\infty$ to $+\infty$
and the system suffers from the Ostrogradsky instability.

We finally mention that the Hamiltonian coincides with the Noether's conserved
quantity corresponding to the invariance of the action under the time translation
when $(\vp,\vq)$ satisfies the canonical equations and constraints.

Hitherto we have focused on the Lagrangian for the third-order EOM. Our analysis also applies to systems with general odd order EOM. We can prove that the Lagrangian for $(2d-1)$th order EOM is given by
\be \label{L-d} L=\sum_{j=1}^{2N}  x^{(d)}_j f_j(\vx, \cdots, \vx^{(d-1)})+g(\vx, \cdots, \vx^{(d-1)}), \ee
with $\det M^{(x^{(d-1)})} \neq 0$.
We can remove ${\ddot \vx}, \cdots, \vx^{(d-1)}$ from the Lagrangian \eqref{L-d} 
by invoking Lagrange multipliers as we have done in the third-order case, 
and proceed to the Hamiltonian formalism with constraints. 
We can then confirm that all the constraints are second class and 
the Hamiltonian is unbounded.

\section{Conclusion}%%%%%%%%%%%%%%%%%%%%%%%%%%%%%%%%%%%%%%%%%
\label{sec:C}
We considered the Lagrangian that yields odd order EOMs in analytical mechanics. 
We explicitly demonstrated how to construct the most general Lagrangian 
for odd order EOMs. 
Using the Hamiltonian formalism for constrained systems, we proved that for this class of theories the Hamiltonian is unbounded. 
Thus, the Ostrogradsky instability persists even in this case.

\acknowledgements{%%%%%%%%%%%%%%%%%%%%%%%%%%%%%%%%%%%%%%%%%
We thank Thierry Sousbie for useful suggestions.
HM was supported in part by Japan Society for the Promotion of Science 
Postdoctoral Fellowships for Research Abroad.
TS was supported in part by Japan Society for the Promotion of Science 
Grant-in-Aid for Scientific Research on Innovative Areas 
No.~25103505 from The Ministry of Education, Culture, Sports, Science and Technology (MEXT),
}

\bibliography{refs}  

\begin{thebibliography}{20}%
\makeatletter
\providecommand \@ifxundefined [1]{%
 \@ifx{#1\undefined}
}%
\providecommand \@ifnum [1]{%
 \ifnum #1\expandafter \@firstoftwo
 \else \expandafter \@secondoftwo
 \fi
}%
\providecommand \@ifx [1]{%
 \ifx #1\expandafter \@firstoftwo
 \else \expandafter \@secondoftwo
 \fi
}%
\providecommand \natexlab [1]{#1}%
\providecommand \enquote  [1]{``#1''}%
\providecommand \bibnamefont  [1]{#1}%
\providecommand \bibfnamefont [1]{#1}%
\providecommand \citenamefont [1]{#1}%
\providecommand \href@noop [0]{\@secondoftwo}%
\providecommand \href [0]{\begingroup \@sanitize@url \@href}%
\providecommand \@href[1]{\@@startlink{#1}\@@href}%
\providecommand \@@href[1]{\endgroup#1\@@endlink}%
\providecommand \@sanitize@url [0]{\catcode `\\12\catcode `\$12\catcode
  `\&12\catcode `\#12\catcode `\^12\catcode `\_12\catcode `\%12\relax}%
\providecommand \@@startlink[1]{}%
\providecommand \@@endlink[0]{}%
\providecommand \url  [0]{\begingroup\@sanitize@url \@url }%
\providecommand \@url [1]{\endgroup\@href {#1}{\urlprefix }}%
\providecommand \urlprefix  [0]{URL }%
\providecommand \Eprint [0]{\href }%
\providecommand \doibase [0]{http://dx.doi.org/}%
\providecommand \selectlanguage [0]{\@gobble}%
\providecommand \bibinfo  [0]{\@secondoftwo}%
\providecommand \bibfield  [0]{\@secondoftwo}%
\providecommand \translation [1]{[#1]}%
\providecommand \BibitemOpen [0]{}%
\providecommand \bibitemStop [0]{}%
\providecommand \bibitemNoStop [0]{.\EOS\space}%
\providecommand \EOS [0]{\spacefactor3000\relax}%
\providecommand \BibitemShut  [1]{\csname bibitem#1\endcsname}%
\let\auto@bib@innerbib\@empty
%</preamble>
\bibitem [{\citenamefont {Woodard}(2007)}]{Woodard:2006nt}%
  \BibitemOpen
  \bibfield  {author} {\bibinfo {author} {\bibfnamefont {R.~P.}\ \bibnamefont
  {Woodard}},\ }\href {\doibase 10.1007/978-3-540-71013-4_14} {\bibfield
  {journal} {\bibinfo  {journal} {Lect.Notes Phys.}\ }\textbf {\bibinfo
  {volume} {720}},\ \bibinfo {pages} {403} (\bibinfo {year} {2007})},\ \Eprint
  {http://arxiv.org/abs/astro-ph/0601672} {arXiv:astro-ph/0601672 [astro-ph]}
  \BibitemShut {NoStop}%
%%CITATION = ASTRO-PH/0601672;%%
\bibitem [{\citenamefont {Smilga}(2009)}]{Smilga:2008pr}%
  \BibitemOpen
  \bibfield  {author} {\bibinfo {author} {\bibfnamefont {A.}~\bibnamefont
  {Smilga}},\ }\href {\doibase 10.3842/Sigma.2009.017} {\bibfield  {journal}
  {\bibinfo  {journal} {SIGMA}\ }\textbf {\bibinfo {volume} {5}},\ \bibinfo
  {pages} {017} (\bibinfo {year} {2009})},\ \Eprint
  {http://arxiv.org/abs/0808.0139} {arXiv:0808.0139 [quant-ph]} \BibitemShut
  {NoStop}%
%%CITATION = ARXIV:0808.0139;%%
\bibitem [{\citenamefont {Ostrogradsky}(1850)}]{Ostrogradsky}%
  \BibitemOpen
  \bibfield  {author} {\bibinfo {author} {\bibfnamefont {M.~V.}\ \bibnamefont
  {Ostrogradsky}},\ }\href@noop {} {\bibfield  {journal} {\bibinfo  {journal}
  {Mem. Acad. St. Petersbourg}\ }\textbf {\bibinfo {volume} {VI 4}},\ \bibinfo
  {pages} {385} (\bibinfo {year} {1850})}\BibitemShut {NoStop}%
\bibitem [{\citenamefont {Pais}\ and\ \citenamefont
  {Uhlenbeck}(1950)}]{Pais:1950za}%
  \BibitemOpen
  \bibfield  {author} {\bibinfo {author} {\bibfnamefont {A.}~\bibnamefont
  {Pais}}\ and\ \bibinfo {author} {\bibfnamefont {G.}~\bibnamefont
  {Uhlenbeck}},\ }\href {\doibase 10.1103/PhysRev.79.145} {\bibfield  {journal}
  {\bibinfo  {journal} {Phys.Rev.}\ }\textbf {\bibinfo {volume} {79}},\
  \bibinfo {pages} {145} (\bibinfo {year} {1950})}\BibitemShut {NoStop}%
%%CITATION = PHRVA,79,145;%%
\bibitem [{\citenamefont {Whittaker.}(1937)}]{Whittaker:1937lec}%
  \BibitemOpen
  \bibfield  {author} {\bibinfo {author} {\bibnamefont {Whittaker.}},\
  }\href@noop {} {\emph {\bibinfo {title} {{A Treatise on the analytical
  dynamics of particles and rigid bodies, fourth edition, p.265}}}}\ (\bibinfo
  {publisher} {Cambridge University Press},\ \bibinfo {address} {London},\
  \bibinfo {year} {1937})\BibitemShut {NoStop}%
\bibitem [{\citenamefont {Nicolis}\ \emph {et~al.}(2009)\citenamefont
  {Nicolis}, \citenamefont {Rattazzi},\ and\ \citenamefont
  {Trincherini}}]{Nicolis:2008in}%
  \BibitemOpen
  \bibfield  {author} {\bibinfo {author} {\bibfnamefont {A.}~\bibnamefont
  {Nicolis}}, \bibinfo {author} {\bibfnamefont {R.}~\bibnamefont {Rattazzi}}, \
  and\ \bibinfo {author} {\bibfnamefont {E.}~\bibnamefont {Trincherini}},\
  }\href {\doibase 10.1103/PhysRevD.79.064036} {\bibfield  {journal} {\bibinfo
  {journal} {Phys.Rev.}\ }\textbf {\bibinfo {volume} {D79}},\ \bibinfo {pages}
  {064036} (\bibinfo {year} {2009})},\ \Eprint {http://arxiv.org/abs/0811.2197}
  {arXiv:0811.2197 [hep-th]} \BibitemShut {NoStop}%
%%CITATION = ARXIV:0811.2197;%%
\bibitem [{\citenamefont {Horndeski}(1974)}]{Horndeski:1974wa}%
  \BibitemOpen
  \bibfield  {author} {\bibinfo {author} {\bibfnamefont {G.~W.}\ \bibnamefont
  {Horndeski}},\ }\href {\doibase 10.1007/BF01807638} {\bibfield  {journal}
  {\bibinfo  {journal} {Int.J.Theor.Phys.}\ }\textbf {\bibinfo {volume} {10}},\
  \bibinfo {pages} {363} (\bibinfo {year} {1974})}\BibitemShut {NoStop}%
%%CITATION = IJTPB,10,363;%%
\bibitem [{\citenamefont {Deffayet}\ \emph {et~al.}(2011)\citenamefont
  {Deffayet}, \citenamefont {Gao}, \citenamefont {Steer},\ and\ \citenamefont
  {Zahariade}}]{Deffayet:2011gz}%
  \BibitemOpen
  \bibfield  {author} {\bibinfo {author} {\bibfnamefont {C.}~\bibnamefont
  {Deffayet}}, \bibinfo {author} {\bibfnamefont {X.}~\bibnamefont {Gao}},
  \bibinfo {author} {\bibfnamefont {D.}~\bibnamefont {Steer}}, \ and\ \bibinfo
  {author} {\bibfnamefont {G.}~\bibnamefont {Zahariade}},\ }\href {\doibase
  10.1103/PhysRevD.84.064039} {\bibfield  {journal} {\bibinfo  {journal}
  {Phys.Rev.}\ }\textbf {\bibinfo {volume} {D84}},\ \bibinfo {pages} {064039}
  (\bibinfo {year} {2011})},\ \Eprint {http://arxiv.org/abs/1103.3260}
  {arXiv:1103.3260 [hep-th]} \BibitemShut {NoStop}%
%%CITATION = ARXIV:1103.3260;%%
\bibitem [{\citenamefont {Kobayashi}\ \emph {et~al.}(2011)\citenamefont
  {Kobayashi}, \citenamefont {Yamaguchi},\ and\ \citenamefont
  {Yokoyama}}]{Kobayashi:2011nu}%
  \BibitemOpen
  \bibfield  {author} {\bibinfo {author} {\bibfnamefont {T.}~\bibnamefont
  {Kobayashi}}, \bibinfo {author} {\bibfnamefont {M.}~\bibnamefont
  {Yamaguchi}}, \ and\ \bibinfo {author} {\bibfnamefont {J.}~\bibnamefont
  {Yokoyama}},\ }\href {\doibase 10.1143/PTP.126.511} {\bibfield  {journal}
  {\bibinfo  {journal} {Prog.Theor.Phys.}\ }\textbf {\bibinfo {volume} {126}},\
  \bibinfo {pages} {511} (\bibinfo {year} {2011})},\ \Eprint
  {http://arxiv.org/abs/1105.5723} {arXiv:1105.5723 [hep-th]} \BibitemShut
  {NoStop}%
%%CITATION = ARXIV:1105.5723;%%
\bibitem [{\citenamefont {Chen}\ \emph {et~al.}(2013)\citenamefont {Chen},
  \citenamefont {Fasiello}, \citenamefont {Lim},\ and\ \citenamefont
  {Tolley}}]{Chen:2012au}%
  \BibitemOpen
  \bibfield  {author} {\bibinfo {author} {\bibfnamefont {T.-j.}\ \bibnamefont
  {Chen}}, \bibinfo {author} {\bibfnamefont {M.}~\bibnamefont {Fasiello}},
  \bibinfo {author} {\bibfnamefont {E.~A.}\ \bibnamefont {Lim}}, \ and\
  \bibinfo {author} {\bibfnamefont {A.~J.}\ \bibnamefont {Tolley}},\ }\href
  {\doibase 10.1088/1475-7516/2013/02/042} {\bibfield  {journal} {\bibinfo
  {journal} {JCAP}\ }\textbf {\bibinfo {volume} {1302}},\ \bibinfo {pages}
  {042} (\bibinfo {year} {2013})},\ \Eprint {http://arxiv.org/abs/1209.0583}
  {arXiv:1209.0583 [hep-th]} \BibitemShut {NoStop}%
%%CITATION = ARXIV:1209.0583;%%
\bibitem [{\citenamefont {Jackiw}\ and\ \citenamefont
  {Pi}(2003)}]{Jackiw:2003pm}%
  \BibitemOpen
  \bibfield  {author} {\bibinfo {author} {\bibfnamefont {R.}~\bibnamefont
  {Jackiw}}\ and\ \bibinfo {author} {\bibfnamefont {S.}~\bibnamefont {Pi}},\
  }\href {\doibase 10.1103/PhysRevD.68.104012} {\bibfield  {journal} {\bibinfo
  {journal} {Phys.Rev.}\ }\textbf {\bibinfo {volume} {D68}},\ \bibinfo {pages}
  {104012} (\bibinfo {year} {2003})},\ \Eprint
  {http://arxiv.org/abs/gr-qc/0308071} {arXiv:gr-qc/0308071 [gr-qc]}
  \BibitemShut {NoStop}%
%%CITATION = GR-QC/0308071;%%
\bibitem [{\citenamefont {Plyushchay}(1988)}]{Plyushchay:1987eu}%
  \BibitemOpen
  \bibfield  {author} {\bibinfo {author} {\bibfnamefont {M.}~\bibnamefont
  {Plyushchay}},\ }\href {\doibase 10.1142/S0217732388001562} {\bibfield
  {journal} {\bibinfo  {journal} {Mod.Phys.Lett.}\ }\textbf {\bibinfo {volume}
  {A3}},\ \bibinfo {pages} {1299} (\bibinfo {year} {1988})}\BibitemShut
  {NoStop}%
%%CITATION = MPLAE,A3,1299;%%
\bibitem [{\citenamefont {Nesterenko}(1989)}]{Nesterenko:1987jt}%
  \BibitemOpen
  \bibfield  {author} {\bibinfo {author} {\bibfnamefont {V.}~\bibnamefont
  {Nesterenko}},\ }\href {\doibase 10.1088/0305-4470/22/10/021} {\bibfield
  {journal} {\bibinfo  {journal} {J.Phys.}\ }\textbf {\bibinfo {volume}
  {A22}},\ \bibinfo {pages} {1673} (\bibinfo {year} {1989})}\BibitemShut
  {NoStop}%
%%CITATION = JPHGB,A22,1673;%%
\bibitem [{\citenamefont {Batlle}\ \emph {et~al.}(1988)\citenamefont {Batlle},
  \citenamefont {Gomis}, \citenamefont {Pons},\ and\ \citenamefont
  {Roman-Roy}}]{Batlle:1987ej}%
  \BibitemOpen
  \bibfield  {author} {\bibinfo {author} {\bibfnamefont {C.}~\bibnamefont
  {Batlle}}, \bibinfo {author} {\bibfnamefont {J.}~\bibnamefont {Gomis}},
  \bibinfo {author} {\bibfnamefont {J.}~\bibnamefont {Pons}}, \ and\ \bibinfo
  {author} {\bibfnamefont {N.}~\bibnamefont {Roman-Roy}},\ }\href {\doibase
  10.1088/0305-4470/21/12/013} {\bibfield  {journal} {\bibinfo  {journal}
  {J.Phys.}\ }\textbf {\bibinfo {volume} {A21}},\ \bibinfo {pages} {2693}
  (\bibinfo {year} {1988})}\BibitemShut {NoStop}%
%%CITATION = JPHGB,A21,2693;%%
\bibitem [{\citenamefont {Saito}\ \emph {et~al.}(1989)\citenamefont {Saito},
  \citenamefont {Sugano}, \citenamefont {Ohta},\ and\ \citenamefont
  {Kimura}}]{Saito:1989bt}%
  \BibitemOpen
  \bibfield  {author} {\bibinfo {author} {\bibfnamefont {Y.}~\bibnamefont
  {Saito}}, \bibinfo {author} {\bibfnamefont {R.}~\bibnamefont {Sugano}},
  \bibinfo {author} {\bibfnamefont {T.}~\bibnamefont {Ohta}}, \ and\ \bibinfo
  {author} {\bibfnamefont {T.}~\bibnamefont {Kimura}},\ }\href {\doibase
  10.1063/1.528331} {\bibfield  {journal} {\bibinfo  {journal} {J.Math.Phys.}\
  }\textbf {\bibinfo {volume} {30}},\ \bibinfo {pages} {1122} (\bibinfo {year}
  {1989})}\BibitemShut {NoStop}%
%%CITATION = JMAPA,30,1122;%%
\bibitem [{\citenamefont {Pons}(1989)}]{Pons:1988tj}%
  \BibitemOpen
  \bibfield  {author} {\bibinfo {author} {\bibfnamefont {J.}~\bibnamefont
  {Pons}},\ }\href {\doibase 10.1007/BF00401583} {\bibfield  {journal}
  {\bibinfo  {journal} {Lett.Math.Phys.}\ }\textbf {\bibinfo {volume} {17}},\
  \bibinfo {pages} {181} (\bibinfo {year} {1989})}\BibitemShut {NoStop}%
%%CITATION = LMPHD,17,181;%%
\bibitem [{\citenamefont {Andrzejewski}\ \emph {et~al.}(2010)\citenamefont
  {Andrzejewski}, \citenamefont {Gonera}, \citenamefont {Machalski},\ and\
  \citenamefont {Maslanka}}]{Andrzejewski:2010kz}%
  \BibitemOpen
  \bibfield  {author} {\bibinfo {author} {\bibfnamefont {K.}~\bibnamefont
  {Andrzejewski}}, \bibinfo {author} {\bibfnamefont {J.}~\bibnamefont
  {Gonera}}, \bibinfo {author} {\bibfnamefont {P.}~\bibnamefont {Machalski}}, \
  and\ \bibinfo {author} {\bibfnamefont {P.}~\bibnamefont {Maslanka}},\ }\href
  {\doibase 10.1103/PhysRevD.82.045008} {\bibfield  {journal} {\bibinfo
  {journal} {Phys.Rev.}\ }\textbf {\bibinfo {volume} {D82}},\ \bibinfo {pages}
  {045008} (\bibinfo {year} {2010})},\ \Eprint {http://arxiv.org/abs/1005.3941}
  {arXiv:1005.3941 [hep-th]} \BibitemShut {NoStop}%
%%CITATION = ARXIV:1005.3941;%%
\bibitem [{\citenamefont {Dirac}(1950)}]{Dirac:1950pj}%
  \BibitemOpen
  \bibfield  {author} {\bibinfo {author} {\bibfnamefont {P.~A.~M.}\
  \bibnamefont {Dirac}},\ }\href {\doibase 10.4153/CJM-1950-012-1} {\bibfield
  {journal} {\bibinfo  {journal} {Can.J.Math.}\ }\textbf {\bibinfo {volume}
  {2}},\ \bibinfo {pages} {129} (\bibinfo {year} {1950})}\BibitemShut {NoStop}%
%%CITATION = CJMAA,2,129;%%
\bibitem [{\citenamefont {Dirac}(1958)}]{Dirac:1958sq}%
  \BibitemOpen
  \bibfield  {author} {\bibinfo {author} {\bibfnamefont {P.~A.~M.}\
  \bibnamefont {Dirac}},\ }\href {\doibase 10.1098/rspa.1958.0141} {\bibfield
  {journal} {\bibinfo  {journal} {Proc.Roy.Soc.Lond.}\ }\textbf {\bibinfo
  {volume} {A246}},\ \bibinfo {pages} {326} (\bibinfo {year}
  {1958})}\BibitemShut {NoStop}%
%%CITATION = PRSLA,A246,326;%%
\bibitem [{\citenamefont {Dirac}(1964)}]{Dirac:1964lec}%
  \BibitemOpen
  \bibfield  {author} {\bibinfo {author} {\bibfnamefont {P.~A.~M.}\
  \bibnamefont {Dirac}},\ }\href@noop {} {\emph {\bibinfo {title} {{Lectures on
  Quantum Mechanics}}}}\ (\bibinfo  {publisher} {Belfer Graduate School of
  Science},\ \bibinfo {address} {New York},\ \bibinfo {year}
  {1964})\BibitemShut {NoStop}%
\end{thebibliography}%
\end{document}